\documentclass{pasj00}

\begin{document}
\SetRunningHead{Deguchi et al.}{SiO Maser Emission from V838 Mon}
\Received{2005/05/11}
\Accepted{2005/07/02 }

\title{Detection of SiO Maser Emission in V838 Mon}

\author{Shuji Deguchi, }
\affil{Nobeyama Radio Observatory, National Astronomical Observatory, \\
and Graduate University for Advanced Studies, 
\\ Minamimaki, Minamisaku, Nagano 384-1305}
\author{Noriyuki Matsunaga,  and Hinako Fukushi}
\affil{Institute of Astronomy, School of Science, The University of Tokyo, 
 \\ Osawa 2-21-1, Mitaka, Tokyo 181-0015}
\author{[PASJ 57 No. 5 (Oct. 25, 2005 issue) in press] }


\maketitle

\begin{abstract}

We report on the detection of 43 GHz SiO maser emission 
in V838 Mon, a prototype of  a new class of eruptive variables, in which a red supergiant
was formed after the nova-like eruption in 2002. The detection of SiO masers
 indicates that the star  formed after the eruption  is indeed a kind of cool mass-losing object with
circumstellar masers.  The measured radial velocity and the intensity of maser emission are
consistent with the object being located at the distance of about 7 kpc from the sun. 
It also suggests that a considerable percentage of SiO masing objects in the Galaxy 
are formed in the same mechanism as that created V838 Mon. 
\end{abstract}

\section{Introduction\label{sec:Intro}}
V838 Monocerotis is a star erupted in the beginning of  January 2002.
After developing an A--F supergiant spectrum at the optical maximum phase
in a few months,  it reveals a cool M-type supergiant spectrum and remains bright in
infrared (\cite{mun02}; \cite{cra03}), appealing to a prototype of a new class of 
eruptive variables (\cite{kim02}). Though a spectacular discovery of a light echo
and  succeeding  observations of the expansion  received  large attentions
(\cite{bon03}; \cite{cra05}), it did not help much to derive an accurate distance 
to this object due to its model dependence (\cite{tyl04}).  
Instead, based on  kinematic and other  information, the distance to this object 
has been estimated to be 8--10 kpc (\cite{mun05}).  

A presence of a  B3V hot companion  (\cite{mun05}) at the post-outburst 
(and likely pre-outburst ; \cite{tyl05}) phase supports a binary-star model of the system.
\citet{sok03} found that thermonuclear models cannot explain the eruption
but a stellar merger can account the outburst luminosity. 
\citet{tyl05} argued from the available observational
data prior to the eruption that the progenitor of V838 Mon was not an evolved red giant,
but  a main sequence star being erupted into an M-supergiant. 

A number of molecular bands such as CO, H$_2$O, and TiO have been detected in absorption in infrared
(\cite{eva03}; \cite{lyn04}).  More recently, \citet{rus05} found variable SiO first-overtone emission at 4 $\mu$m,
which indicates a characteristic of  cool supergiant.
A search for circumstellar molecular emission at radio wavelengths 
(in  the SiO $J=2$--1 $v=1$ line, and the $^{12}$CO $J=1$--0, 2--1, and 3--2 lines) 
gave negative results (\cite{rus03}). 

In this paper, we report on the SiO maser detection toward the unusual eruptive variable,
V838 Mon.  The observations made in a two-month separation (February and April, 2005)
indicate that the  SiO maser intensity is increasing at current phase.
The detection of SiO masers in this object implies that some percentages of  stellar maser sources,
which have been considered to be mass losing stars at the Asymptotic Giant Branch (AGB) 
phase (or occasionally post-AGB phase),
can be created by the same mechanism as that created V838 Mon.
We discuss on the implication of this result in section 3.

\section{Observations and Results\label{sec:Obs}}

The first observation of V838 Mon with Nobeyama 45-m telescope was made on 2005 February 23
in the SiO maser lines ($J=1$--0,  $v=1$ and 2) at 43.122 and 42.821 GHz, respectively.
The half-power full beam width (HPFBW) was about 40$''$ at 43 GHz. 
We used a cooled SIS-mixer receiver ($T_{\rm sys} \sim$ 180 -- 250 K)
and acousto-optical spectrometers with high (40 kHz; AOS-H) and
low (250 kHz; AOS-W) resolutions having 2048 channels each.
The spectrometer arrays
covered  velocity ranges of $\pm 390$ km s$^{-1}$ and  $\pm 800$ km s$^{-1}$
in AOS-H and AOS-W with effective velocity resolution of 0.3 and  1.8 
km s$^{-1}$ per binned channel,  respectively.
The conversion factor of the antenna temperature ($\equiv T_a^*$) to the flux density was
$\sim 2.9$ Jy K$^{-1}$. The detail of observations with this system had been described elsewhere
(for example, \cite{deg00}).  

\begin{figure*}
\begin{center}
\FigureFile(170mm,100mm){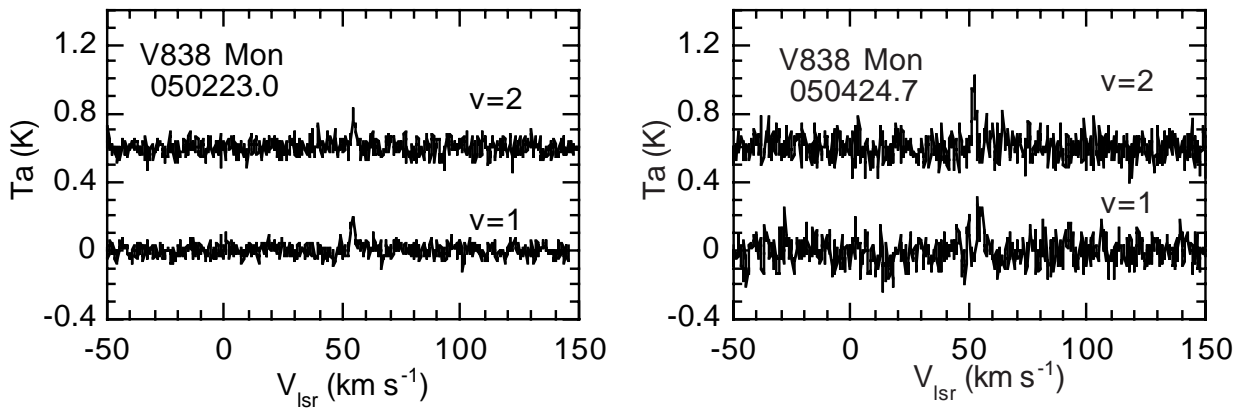}
\end{center}
\caption{Spectra of the SiO $J=1$--$0, v=2$ (top) and $v=1$ line (bottom)
in V838 Mon. The observation dates are 2005 February 23 (left panel) and 
2005 April  24 (right panel).
\label{fig:1}
}
\end{figure*}

We detected SiO maser emission in both $J=1$--0 $v=1$ and 2 transitions  
at $V_{LSR}\sim 54$ km s$^{-1}$ 
in V838 Mon in about 11 minutes on-source integration time.
The spectra taken in AOS-H are shown in figure \ref{fig:1}.
The detections were confirmed in AOS-W spectra in all cases.
The line properties
are summarized in table \ref{tab:1}, which contains the source name,
observed positions, observed transitions, 
radial velocities ($V_{\rm LSR}$), peak antenna temperatures ($T_{\rm a}$),
integrated intensities,
and rms noise levels for both transitions from the AOS-H spectra.
The MSX 6C catalog (\cite{ega03}) gives no point source detected within 10$'$ of V838 Mon 
in all of the observed bands, although the MSX imager shows a slight enhancement of 
emission in A (8.3 $\mu$m) band at the position of IRAS 07015$-$0346 
(=V838 Mon; $F_{12}=0.25$ Jy; see \cite{van04}).
The 2MASS images (\cite{cul03}) do not show any contaminating bright red objects within 1$'$ of V838 Mon.
Therefore, there is no chance of contamination by other objects emitting SiO masers
within a telescope beam.

A short follow-up observation with the same telescope was made on 2005 April 24  
for  V838 Mon in the SiO $J=1$--0, $v=1$ and 2, and 
$J=2$--1, $v=1$ (86.243 GHz) lines.
The AOS-H spectrometer arrays 
covered the radial velocity range between   $\pm 75$   km s$^{-1}$  in the 86 GHz SiO maser  observation.
The system temperature was approximately 240 K
at 86 GHz.  The conversion factor of the antenna temperature to the flux density
was  inferred to be approximately 4.4 Jy/K at  86 GHz. The $J=1$--0 $v=1$ and 2 maser lines around 43 GHz 
were found to become slightly stronger than before, but the $J=2$--1 $v=1$ line was not detected.

Figure 1 and table \ref{tab:1} indicate that the SiO maser intensity  in V838 Mon apparently increased 
with time in a two-month span.
 Though the calibration of the line flux involved uncertainty of about $\pm$  20\% in the NRO  45 m telescope system
 (for example, see \cite{kamh05}), and  the noise level was high in the
 2005 April  observation due to time restriction, the maser intensity  in 2005 April seems to 
increase by about 50 \% compared with the intensity in 2005 February.  It is consistent with the fact that  
the 4 $\mu$m SiO  first-overtone emission was first developed in a periphery of the extended atmosphere 
during 2002--2003 (\cite{rus05}),  and after a year or two  the outflowing gas made SiO masers at  the outer envelope. 
An outflow velocity of 20 km s$^{-1}$ gives
a crossing length of about $6 \times 10^{13}$ cm over a year,  when the gas  reaches to the radius of $\sim 1.6 \times 10^{14}$  cm
if it started at the photospheric radius  of  $1.0 \times 10^{14}$ cm, which was measured in the near-IR  interferometer
at 2.2 $\mu$m (\cite{lan05}). 

In addition,  we observed the same type of eruptive variable, V4332 Sgr (\cite{mar99}; \cite{ban04}) 
on 2005 March 9 and 10 by the SiO $J=1$--0 $v=1$ and 2, 
and added the negative results in table \ref{tab:1}. 
This star, V4332 Sgr, was erupted in 1994 February,
developing early M-type spectrum involving TiO bands (\cite{tyl05b}). 
The distance was inferred to be a few kpc away from the Sun. 
No maser was detected in this star. The 2MASS database shows a faint red star with $K=10.99$ 
and $H-K=0.61$  at this position. 

\begin{table*}
  \caption{Observed  line intensities }\label{tab:1}
 \begin{center}
  \begin{tabular}{lcccccccr}
  \hline\hline
Object & RA(J2000)$^{*}$ & Dec(J2000)$^{*}$  & Transition & $V_{\rm LSR}$ & 
Peak $T_{\rm a}$ & Integ. int. &  rms  & obs.date \\
  &  h ~ m ~ s & $^\circ ~~' ~~ ''$  &  mol, $J_u$--$J_l$, $v$ & (km~s$^{-1})$ & (K) &
(K~km~s$^{-1}$) & (K)  & yymmdd.d \\
\hline
V838 Mon & $07~04~04.85$ & $-03~50~51.1$
 & SiO 1--0, ~1& 54.6 & 0.197 & 0.577 & 0.034 & 050223.0 \\ 
   & &  &  SiO 1--0, ~2 &  54.2 & 0.251 & 0.396 & 0.042  & 050223.0   \\ 
   &  &  & SiO 1--0, ~1& 53.5 & 0.330 & 0.813 & 0.070 & 050424.7 \\ 
   & & &  SiO 1--0, ~2 &  54.1 & 0.419 & 0.681 & 0.065  & 050424.7   \\ 
   & & &  SiO 2--1, ~1 &   ...  &  ...  & ... & 0.034  & 050424.7   \\ 
V4332 Sgr & $18~50~36.70$ & $-21~23~29.6$
  &  SiO 1--0, ~1& ... & ... & ... & 0.052  & 050310.3 \\ 
  & & & SiO 1--0, ~2 & ... & ... & ... & 0.068 & 050310.3  \\
   \hline
\multicolumn{9}{l}{$^{*}$ : Positions were taken from the SIMBAD database [originally from 
\citet{bro02} for V838 Mon,}\\
\multicolumn{9}{l}{ and  \citet{dow01} for V4332 Sgr].}   
  \end{tabular}
  \end{center}
\end{table*}


\section{Discussions\label{sec:disc}}

\subsection{Radial Velocity, Maser Intensity,  and Kinematic Distance of V838 Mon.\label{sec:velocity}}
The radial velocity of SiO maser emission is
known to coincide with that of the central star within a few km s$^{-1}$ \citep{jew91}.
It directly indicates the velocity of a cool M-supergiant, which was formed after eruption.
The relatively narrow ($\sim$ 5 km s$^{-1}$) widths of both maser lines
suggest that the turbulence in the outflowing envelope of the M supergiant is
relatively mild. The SiO maser lines are normally formed at a few stellar radii
of the photosphere for M-type stars ($\sim 10^{14}$ cm; \cite{cot04}),
but  for supergiants, VLBA observations indicate that it is slightly outer 
part of the envelope (\cite{miy03}). The SiO  maser  line profiles of  V838 Mon do not show
any indication of  the broad line width which is common for supergiants(\cite{cer97}).
This is partly due to  the low signal-to-noise ratio of the detected lines,
in which the profile is not enough to reveal the weak broad pedestal feature often seen in supergiants.

The radial velocity of V838 Mon was not accurately known though it has been 
estimated from P-Cygni-type optical spectra (\cite{kip04}; \cite{tyl05}) as $V_{\rm helio}=$ 55--65 km s$^{-1}$
(corresponding to $V_{\rm LSR}=$42--52 km s$^{-1}$). The SiO radial velocity
found in the present paper, $V_{\rm LSR}$=54 km s$^{-1}$,  coincides with the high end 
of the optical velocities, 
establishing  the accurate stellar radial velocity.  \citet{tyl05} listed the radial velocities 
of interstellar clouds 
within $\sim 1.5 ^{\circ}$ from V838 Mon. From their table 3,
we find that two molecular clouds, G217.7+2.4 and G218.7+1.8 (=IC 466 or S288),
which have the highest radial velocities in the table ($V_{\rm LSR}=$54.8 and 56.8 km s$^{-1}$), have
the similar radial velocities as that of  V838 Mon. The kinematic distances
to these clouds were evaluated to be 7.02 and 7.17 kpc, respectively
(\cite{woo89}).
\citet{jia96} estimated the kinematic distances of SiO maser sources
in this direction using the rotation curve derived by \citet{bur88}. 
Their figure 8 gives the kinematic distance of
about 7 kpc for  $V_{\rm LSR}=$54 km s$^{-1}$, suggesting that
V838 Mon is located at the similar distance with these molecular clouds
(if V838 Mon is a disk population). 

The intensity of SiO maser lines,  0.6 -- 1.2  Jy, is comparable with
the intensities of masers found in the galactic center  and  bulge (e.g., see \cite{deg02}; \cite{deg04a}). 
Therefore, this fact also supports the large distance ($\sim 7$ kpc) of V838 Mon,
provided that the M-supergiant of V838 Mon emits a similar intrinsic maser  flux as the bulge SiO maser sources.
However, because only  maximum (upper limit) of SiO maser line intensity relative to
the IRAS 12 $\mu$m flux density (i.e., normalized by the luminosity and the distance of the object) 
is meaningful  in general  (because of time variation
of maser intensity; \cite{jew91}), we cannot completely deny the possibility of smaller distance
merely based on the line fluxes.   
The measured flux density is nearly equal to the value expected from
the maximum photon fluxes of the usual SiO sources
at the distance of 7--8 kpc, indicating that the envelope of an M-supergiant 
in V838 Mon radiates  SiO masers at a nearly maximum flux  among masing objects.

\subsection{Eruptive Formation of Maser Stars \label{sec:EFMS}}
Though the interstellar matter revealed in the light echo of V838 Mon 
and a diffuse middle infrared emission found in MSX map
 were claimed to be a past activity of
AGB phase of the progenitor  of V838 Mon (\cite{van04}),   
a current understanding of photometric data of the progenitor star
and modeling of light echo  seem to conflict with
the RGB/AGB/post-AGB hypothesis of the progenitor (\cite{tyl05}).
Rather, considering various possibilities, \citet{tyl05} argued that the progenitor
of  V838 Mon was a binary system of two main-sequence stars
such as B1.5V and B3V, and that the interstellar material, which was brightened by light echo,
does not originate from V838 Mon.
  
It is interesting to consider a scenario of the binary evolution, 
which produces an SiO maser star (late M supergiant) by an
eruptive event.  From the fact that we know already two examples
of the eruptive formation of an M supergiant in the Galaxy: V838 Mon and V4332 Sgr (e.g., \cite{ban04}).
We estimate that the formation rate of this type of 
new class of nova events creating an M-supergiant after eruption
is roughly one in every 10 years in the Galaxy.  Because we could not detect
any SiO maser emission in V4332 Sgr in the present paper,
we assume that one of these two events creates an SiO maser star.
It is  hard to estimate the life time of maser emitting  
phase in V838 Mon. However, let us assume a rather  buoyant value,
2000 years.  In this case, we should observe 100 such maser stars
of this kind in the Galaxy at any epoch of time. 
Furthermore,  we know approximately about 1500 SiO maser emitting objects
in the Galaxy (\cite{deg04b}) and  about 1000 OH/IR sources (\cite{sev01}).
Therefore, about 5--10\% of these objects
have the eruptive origin.

This is, of course, a very rough estimate, merely giving an order of magnitudes for such percentage.
Furthermore,  the galactic nova rate is difficult to estimate because of the patchy interstellar extinction
(\cite{sha97}).   The largest uncertainty in the previous estimate seems to be involved
in the time span of SiO maser
emitting phase in V838 Mon at present.  Because the SiO masers are emitted at
just a few stellar radii of the photosphere, it terminates within a few years when the mass loss
ceases. Because once an AGB star is created (as a single star), its life time at the AGB phase 
would be longer than $10^5$ years (depending on the mass). 
Therefore, it is not expected that the SiO maser terminates quickly
unless the formed AGB-like structure is quite unstable and transient.  
It is quite interesting to know how long the SiO masers are detectable in V838 Mon 
and  to know whether  or not H$_2$O and OH masers follow in future. 

We discussed the eruptive origins of maser stars based on the view point 
of a  binary merger model of V838 Mon for simplicity. However, it should be noted 
that the above arguments are equally applied for an alternative scenario, i.e., the post-AGB progenitor
being born again as an AGB star (\cite{law03}), though the life span of the maser emitting phase 
must significantly be altered in such a scenario.

Spectroscopic observations suggested that V838 Mon is slightly metal deficient
except with enhanced abundances of s-process elements (\cite{kip04}; \cite{kam05}).
Among metallic species, \citet{kip04} gave the Si abundance as 1/10 of the solar value,
which does not seem to be consistent with the later finding of rich SiO infrared band emission
(\cite{rus05}) and the SiO maser detection in the present paper. These facts may imply 
extreme difficulty of the abundance analysis from transient optical spectra.
Alternative evidence of binary mergers in  slightly metal poor environments  was recently found 
in bulge globular clusters (\cite{mat05});  some SiO maser sources toward globulars are likely 
to be cluster members. Because luminosities of these objects
slightly exceed the AGB luminosity limit of the low-mass objects expected 
in the aged globular clusters, these stars must be the objects created by binary mergers.
In accounting these SiO maser sources in globulars, observations of maser sources do not necessarily 
constrain the mechanism whether it is a sudden, eruptive formation of an M-supergiant or  it is
 a merger event  in which a merged star evolved into the AGB phase in a certain period later.
 Statistics of SiO maser sources and blue stragglers in globulars may provide an answer 
 to this question.   

\section{Conclusion}
SiO maser emission from the eruptive variable, V838 Mon, was detected
with the Nobeyama 45m telescope, confirming formation of  a mass-losing M supergiant
after nova eruption. The obtained radial velocity of masers, $V_{\rm LSR}\sim 54$ km s$^{-1}$,
gives the first reliable evaluation of the radial velocity of  this star, suggesting the
kinematic distance of about 7 kpc for this object.
If the SiO maser phenomenon in this star is persistent in a certain length,
a considerable percentage of the SiO maser stars in the Galaxy may originate though 
the same mechanism, i.e.,  eruptive formation of  maser stars from merged binaries.   
This observation  approves evidence of the SiO maser phenomenon occurring
 in diversely different types of objects.
 
\

The authors thank  Drs. I. Yamamura, T. Fujii, Y. Nakada, T. Tanabe,  and Y. Ita 
for discussions and  encouragements on this work.  This research 
makes use of the SIMBAD database operated at CDS, Strasbourg, France,
 as well as data products from
the Two Micron All Sky Survey, which is a joint project of the University of
Massachusetts and the Infrared Processing and Analysis Center/California
Institute of Technology, funded by the National Aeronautics and Space
Administration and the National Science Foundation. 


\end{document}